# Pareto's 80/20 Rule and the Gaussian Distribution

Katsuaki Tanabe*

*Department of Chemical Engineering, Kyoto University, Nishikyo, Kyoto 615-8510, Japan*

*E-mail: tanabe@cheme.kyoto-u.ac.jp

## Abstract

The statistical state for the empirical Pareto's 80/20 rule has been found to correspond to a normal or Gaussian distribution with a standard deviation that is twice the mean. This finding represents large characteristic variations in our society and nature. In this distribution, the rule can be also referred to as, for example, the 25/5, 45/10, 60/15, or 90/25 rule. In addition, our result suggests the existence of implicit negative contributors.

## Introduction

Pareto's 80/20 rule states that roughly 80% of all effects stem from 20% of all causes for many events, which conceptually contrasts the contribution of the vital few with that of the trivial many.[1–3] This rule has been applied in a variety of fields, such as economics,[4–7] biology,[8,9] ethology,[10,11] and civil engineering,[12,13] where its validity and usefulness have been demonstrated. Mathematically, the 80/20 rule is often interpreted as an instance of the Pareto distribution.[14–16] However, the power law of the Pareto distribution is originally a model intended to represent the probability of a variable exceeding a certain threshold value, and is also known as the survival or tail function. Meanwhile, many distributions pertaining to events in our society and nature, to which the 80/20 rule is often applied, more commonly follow the normal or Gaussian distribution. In other words, it may be more intuitive to assume a distribution with a peak around the average or mean in order to discuss the 80/20 rule, rather than the monotonic Pareto distribution. Therefore, in this short note we present an analysis of Pareto's 80/20 rule based on the Gaussian distribution.

## Theory and Calculation Methods

The normal or Gaussian probability distribution $f(x)$ based on the central limit theorem is described as

$$f(x) = \frac{1}{\sqrt{2\pi}\sigma} \exp\left\{ -\frac{(x-\mu)^2}{2\sigma^2} \right\} \qquad (1),$$

where $\sigma$ and $\mu$ are the standard deviation and the mean, respectively. To analyze and discuss Pareto's 80/20 rule, we define the cause and effect integrated fractions, respectively $I_{cause}$ and $I_{effect}$, as

$$I_{cause} \equiv \frac{\int_{\mu+X}^{\infty} f(x)dx}{\int_{-\infty}^{\infty} f(x)dx} = \int_{\mu+X}^{\infty} f(x)dx \qquad (2),$$

$$I_{effect} \equiv \frac{\int_{\mu+X}^{\infty} xf(x)dx}{\int_{-\infty}^{\infty} xf(x)dx} = \frac{1}{\mu}\int_{\mu+X}^{\infty} xf(x)dx \qquad (3).$$

In popular folklore, the 80/20 rule includes such claims as that 20% of a population own 80% of the wealth, that 20% of the books in a library account for 80% of the circulation, that 20% of a business customers bring in 80% of its revenue, that 20% of all software features account for 80% of all software use, and so on. The quantity $I_{cause}$ is a proportion of the population; the quantity $I_{effect}$ is the proportion of the total income that they get. Or $I_{cause}$ is a proportion of the population of books in a library and $I_{effect}$ is the corresponding proportion of all circulation. And so on. Here, we note that

$$\int_{-\infty}^{\infty} f(x)dx = 1 \qquad (4),$$

$$\int_{-\infty}^{\infty} xf(x)dx = \mu \qquad (5),$$

for the definitions of $f(x)$ and $\mu$. Figure 1 depicts an example of the Gaussian distribution, along with its corresponding $I_{cause}$ and $I_{effect}$, for the case that $\sigma = 1$ and $\mu = 2$. $X$ denotes the threshold deviation from $\mu$ for defining $I_{cause}$ and $I_{effect}$. As a common characteristic of the Gaussian distribution, it is well known, for example, that $I_{cause}$ = 0.16, 0.023, and 0.0014 for $X = \sigma$, $2\sigma$, and $3\sigma$, respectively. In Fig. 1(b), note that $I_{effect}$ is not exactly the integrated area in the graph, but rather that divided by $\mu$, as shown in Eq. 3. Regarding the similarity of distributions, same $\sigma/\mu$ values yield the same $I_{effect}$-$I_{cause}$ relation, regardless of the individual absolute values of $\sigma$ and $\mu$; therefore, we can conduct investigations based just on the ratio of $\sigma/\mu$. In other words, the $I_{effect}$-$I_{cause}$ relation is only a function of the ratio $\sigma/\mu$ rather than the individual values of $\sigma$ and $\mu$. We calculate $I_{effect}$ and $I_{cause}$ with various values of $\sigma/\mu$ and $X$.

## Results and Discussion

Figure 2 shows the relationship calculated between $I_{cause}$ and $I_{effect}$ for varied values of $X$ and $\sigma/\mu$. We see that the point ($I_{cause}$, $I_{effect}$) = (0.2, 0.8) lies roughly in the curve for $\sigma/\mu = 2$. Importantly, this curve ($\sigma/\mu \sim 2$) generalizes the 80/20 rule. Interestingly, this result implies that the statistical state for the 80/20 rule (i.e., the state where the rule holds) corresponds to a distribution with a standard deviation that is twice the mean. This result also indicates that human society lies in such a state. Deducing inversely from the empirical Pareto's 80/20 rule, we find that our society and the nature are highly dispersive. Incidentally, it should be noted that the region above 100% in the effect for the corresponding curve plotted in Fig. 2 is not a mathematical artifact but appears in reality; data for example shows that the vital component of the customers can often provide over 100% of the total profit to a company,[17] in conjunction with the existence of negative contributors to be discussed in the following. Figure 3 shows the Gaussian distribution for $\sigma/\mu = 2$, representing the state of Pareto's 80/20 rule. As seen in Fig. 3(b), our result might also suggest the existence of implicit negative factors (i.e., some causes can provide less-than-zero contribution). This phenomenon is actually observed in the society, for example, in the form that some customers rather bring financial losses to companies.[17] In other words, our result presented as Fig. 3 provides a quantitative reasoning of the existence of such negative contributors. Figure 4 plots $I_{cause}$ and $I_{effect}$ for the Gaussian distribution for $\sigma/\mu = 2$, which represents the state for Pareto's 80/20 rule, depending on $X$. From these $I_{cause}$ and $I_{effect}$ curves, what we understand is that, similar to the 80/20 rule, we can also derive a 25/5 rule, i.e., we can deduce that 20% of the effects are caused by 5% of the causes in a Gaussian distribution. We can also further create new other rules in similar manners. We thus recognize from the plot that the 80/20 rule can also be read as the "25/5 rule" ($X = 1.7\sigma$), "45/10 rule" ($1.3\sigma$), "60/15 rule" ($1.1\sigma$), "90/25 rule" ($0.67\sigma$), and so forth. As touched in the *Introduction* section, the Pareto distribution (*Type I* ) $f_P(x)$ is described as

$$f_P(x) = \alpha {x_{\min}}^{\alpha} \frac{1}{x^{\alpha+1}} \qquad (6),$$

where $\alpha$ (> 0) is a shape parameter called the *Pareto index*, and $x_{\min}$ (> 0) is the minimum possible value of $x$. Incidentally, this Pareto distribution is plainly not realistic in many fields, e.g., implying that nobody has an income less than $x_{\min}$. Note that

$$\int_{x_{\min}}^{\infty} f_P(x)dx = 1 \qquad (7).$$

Then $I_{cause}$, or the probability that $x$ is greater than some value $A$ ($> x_{\min}$) is, for $f_P(x)$ is:

$$I_{cause} \equiv \frac{\int_A^{\infty} f_P(x)dx}{\int_{x_{\min}}^{\infty} f_P(x)dx} = \int_A^{\infty} f_P(x)dx = \left(\frac{x_{\min}}{A}\right)^{\alpha} \qquad (8),$$

and $I_{effect}$ is:

$$I_{effect} \equiv \frac{\int_A^{\infty} x f_P(x)dx}{\int_{x_{\min}}^{\infty} x f_P(x)dx} = \left(\frac{x_{\min}}{A}\right)^{\alpha-1} \qquad (9).$$

We assumed $\alpha > 1$. Otherwise, $I_{effect} = 1$. Solving Eqs. 8 and 9 for $\alpha$ when $I_{cause} = 0.2$ and $I_{effect} = 0.8$, we obtain $\alpha = \log_4 5 \sim 1.16$. This is why it is claimed that the 80/20 rule holds when $\alpha = \log_4 5$ based on the Pareto distribution. However, in the case of this power-law logic, an iterated 80/20 rule necessarily comes along as follows. From Eqs. 8 and 9,

$$\frac{\log I_{effect}}{\log I_{cause}} = \frac{\alpha - 1}{\alpha} = \frac{\log_4 5 - 1}{\log_4 5} = \frac{\log 0.8}{\log 0.2} \qquad (10).$$

Therefore, when $I_{cause} = 0.2^n$, $I_{effect} = 0.8^n$. The number $n$ of iterations needs not be an integer. It thus necessarily follows the "64/4 rule" ($n$ = 2), "51.2/0.8 rule" ($n$ = 3),

"40.96/0.16 rule" ($n$ = 4), and so forth. For comparison, Fig. 5 shows the relationship between $I_{cause}$ and $I_{effect}$ for this iterated 80/20 rule based on the Pareto distribution, plotted along with the curve for the Gaussian distribution for $\sigma/\mu = 2$. The $I_{effect}$-$I_{cause}$ relation based on the Pareto distribution, particularly in the low-fraction region, seems too drastic for the real world, e.g., a half of the total wealth may not be occupied by < 1% of the people. In contrast, our series of $I_{effect}$-to-$I_{cause}$ ratio (25/5, 45/10, 60/15, etc.) based on the Gaussian distribution may sound more realistic in many cases.

## Conclusions

In this short note, we have examined the empirical Pareto's 80/20 rule from the perspective of the normal or Gaussian probability distribution. We found that the 80/20 rule represents the case when the standard deviation is twice the mean in the Gaussian distribution. This result implies a high diversity of characteristics in society and nature. Our result might also suggest the existence of implicit negative factors.

## Acknowledgements

We thank an anonymous reviewer for her/his essential insight on the power law.

## Figure Captions

**Fig. 1.** Example of Gaussian distribution (a) $f(x)$ and (b) $x$f($x$) along with its corresponding $I_{cause}$ and $I_{effect}$ regions for $\sigma = 1$, $\mu = 2$.

**Fig. 2.** Relationship between $I_{cause}$ and $I_{effect}$ under various $X$ and $\sigma/\mu$ values. The curve for $\sigma/\mu = 2$ contains the point ($I_{cause} = 0.2$, $I_{effect} = 0.8$), which represents the state for Pareto's 80/20 rule.

**Fig. 3.** Gaussian distribution with $\sigma/\mu = 2$, representing the state for Pareto's 80/20 rule. (For this figure we set $\sigma = 2$, $\mu = 1$, but individual values do not matter, for the shape is the same as long as $\sigma/\mu$ is identical.)

**Fig. 4.** $I_{cause}$ and $I_{effect}$ against $X$ for the Gaussian distribution of $\sigma/\mu = 2$, which represents the state for Pareto's 80/20 rule.

**Fig. 5.** Relationship between $I_{cause}$ and $I_{effect}$ for the Gaussian distribution ($\sigma/\mu = 2$) and the Pareto distribution.

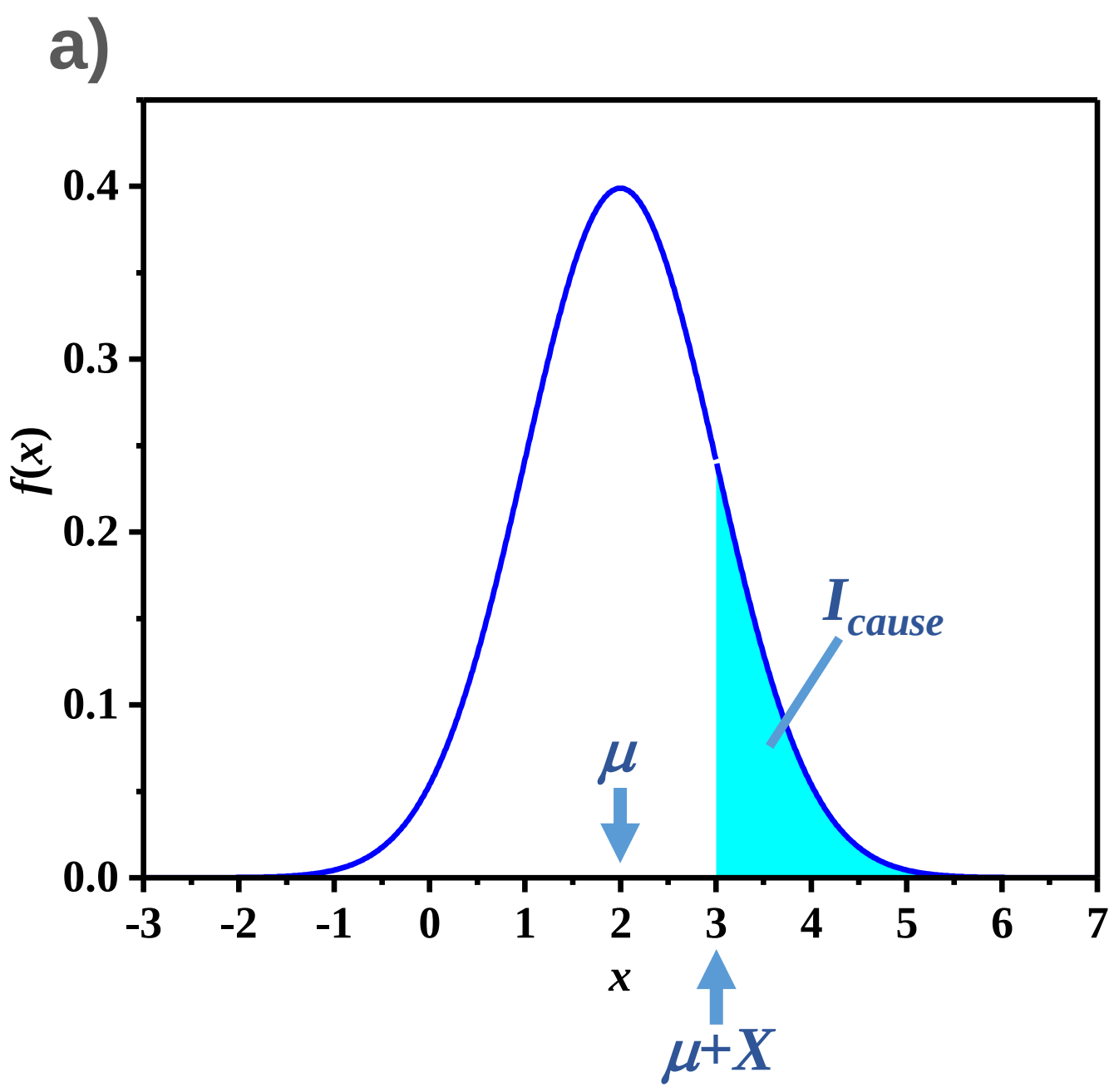

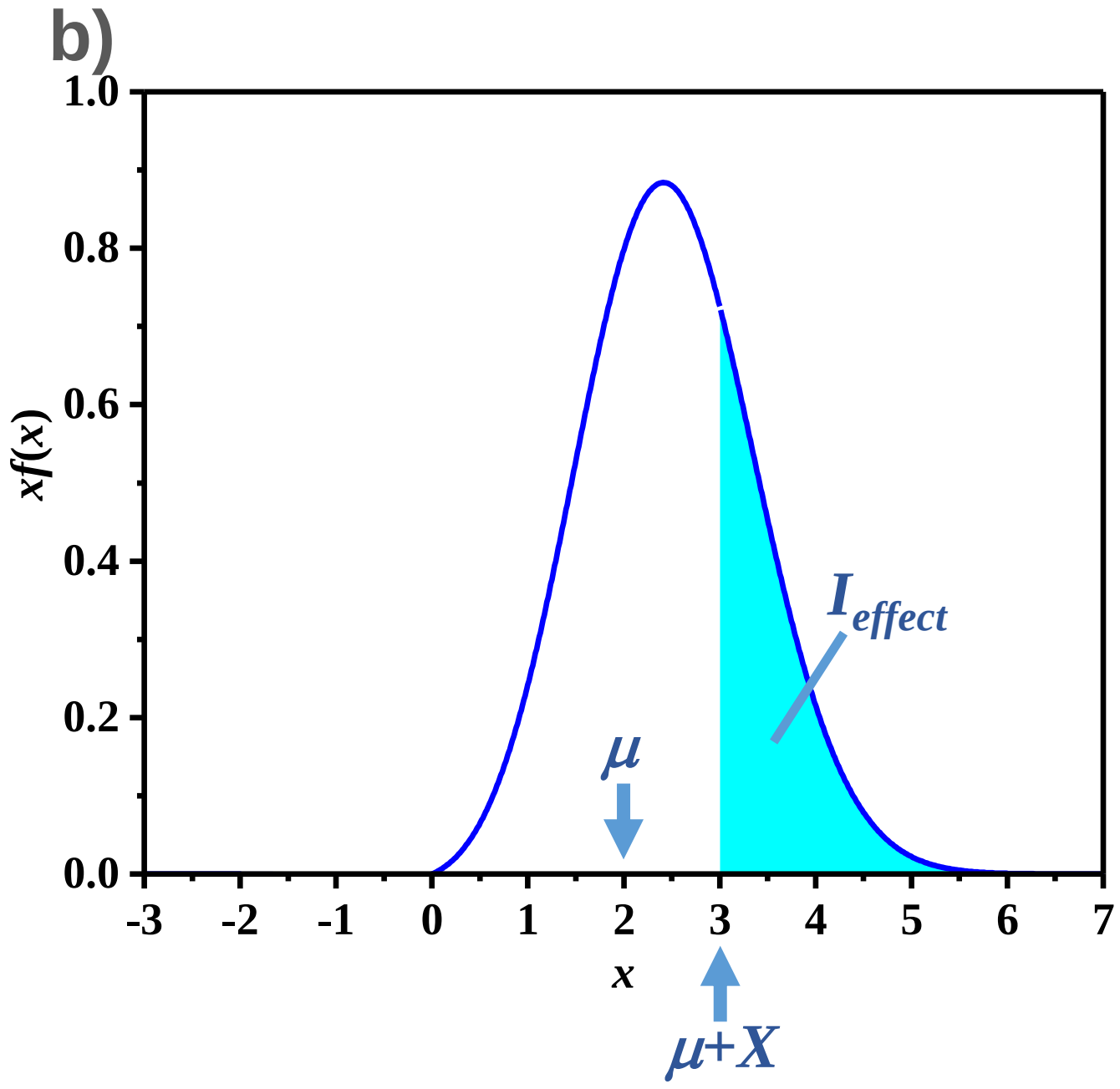

Fig. 1.

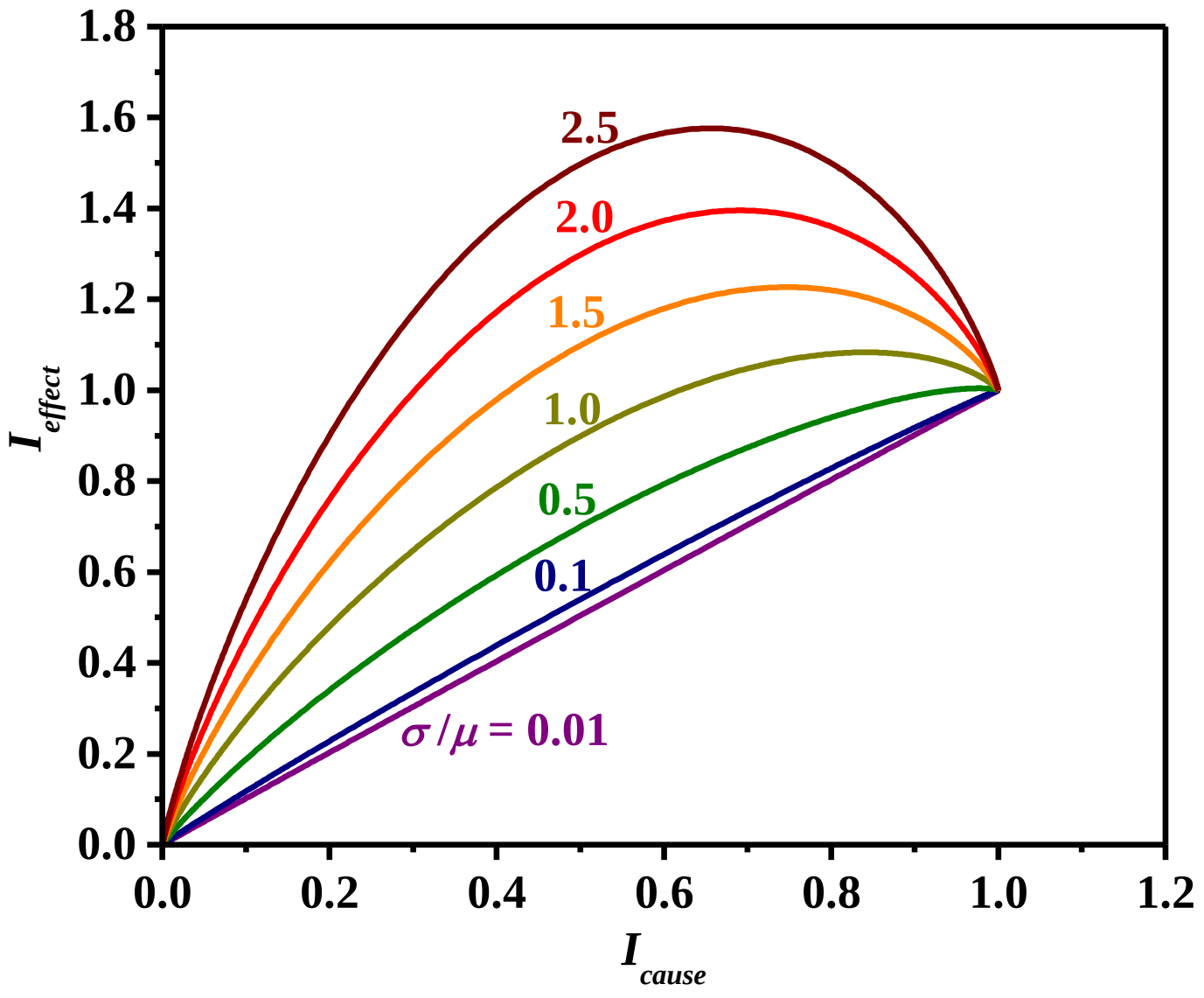

Fig. 2.

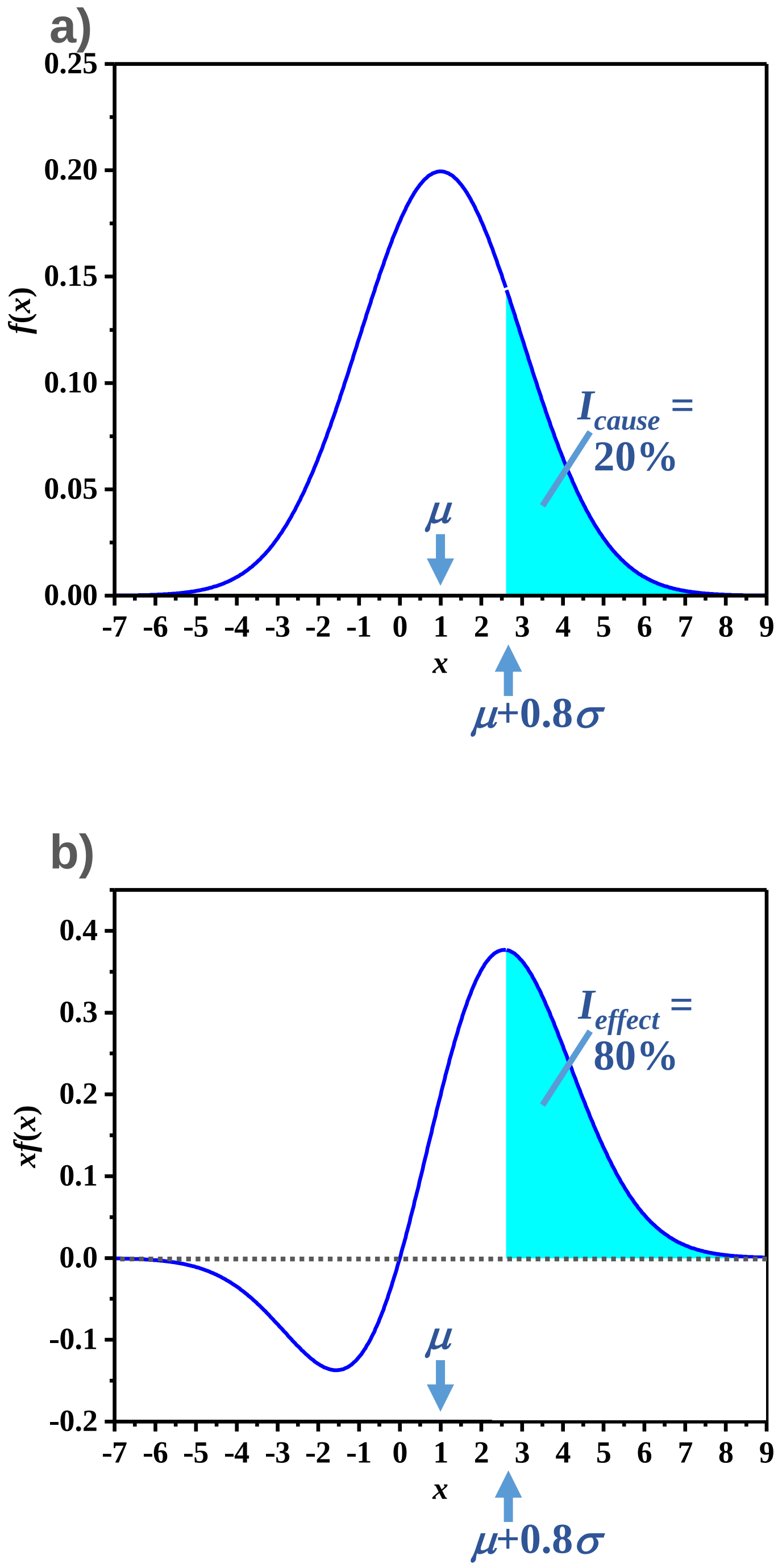
a)
0.25
0.20
0.15
0.10
0.05
0.00
f(x)
-7 -6 -5 -4 -3 -2 -1 0 1 2 3 4 5 6 7 8 9
x
μ
μ+0.8σ
I cause =
20%
b)
0.4
0.3
0.2
0.1
0.0
-0.1
-0.2
xf(x)
-7 -6 -5 -4 -3 -2 -1 0 1 2 3 4 5 6 7 8 9
x
μ
μ+0.8σ
I effect =
80%

Fig. 3.

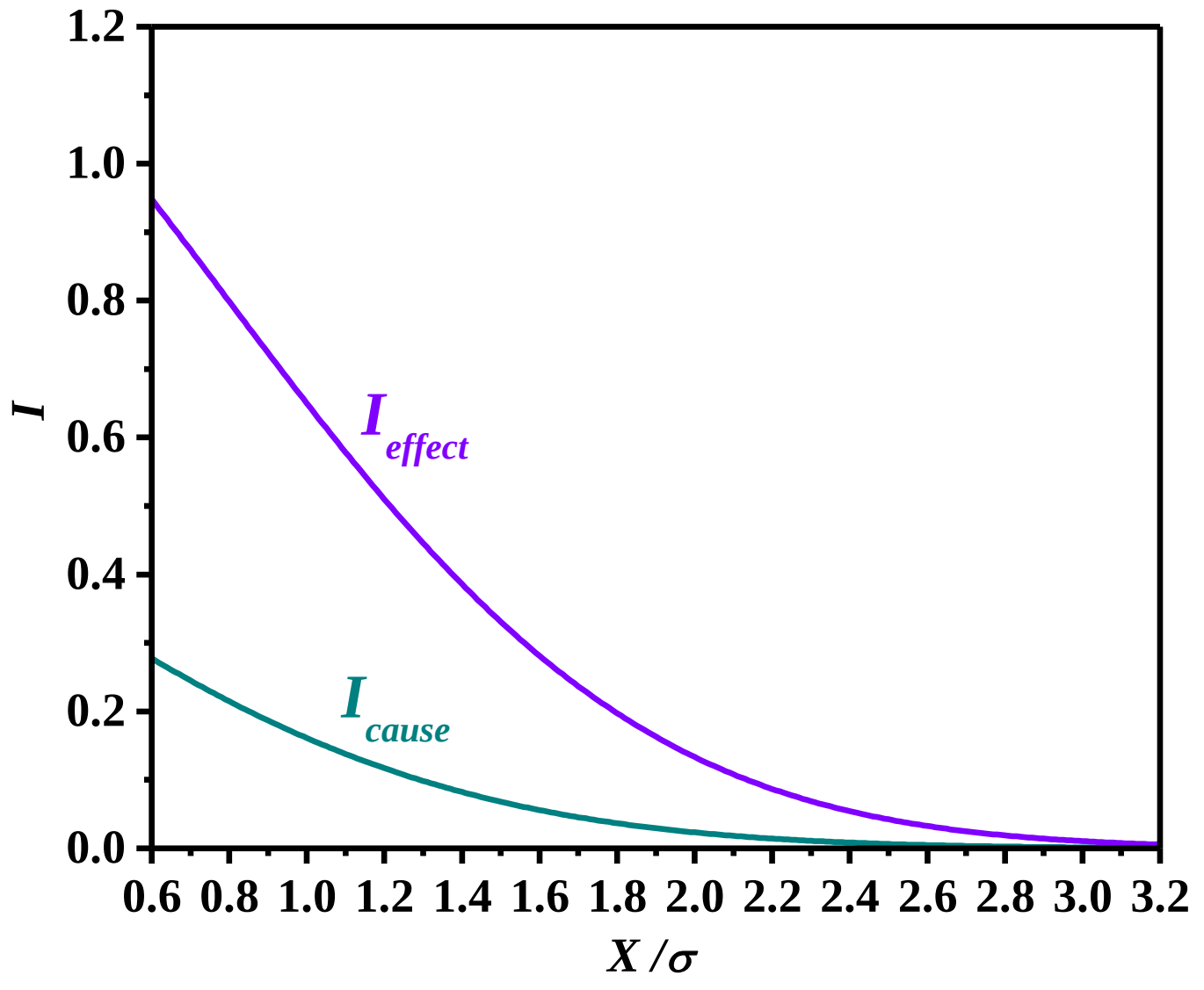

1.2
1.0
0.8
0.6
0.4
0.2
0.0
I
I effect
I cause
0.6 0.8 1.0 1.2 1.4 1.6 1.8 2.0 2.2 2.4 2.6 2.8 3.0 3.2
X /σ

Fig. 4.

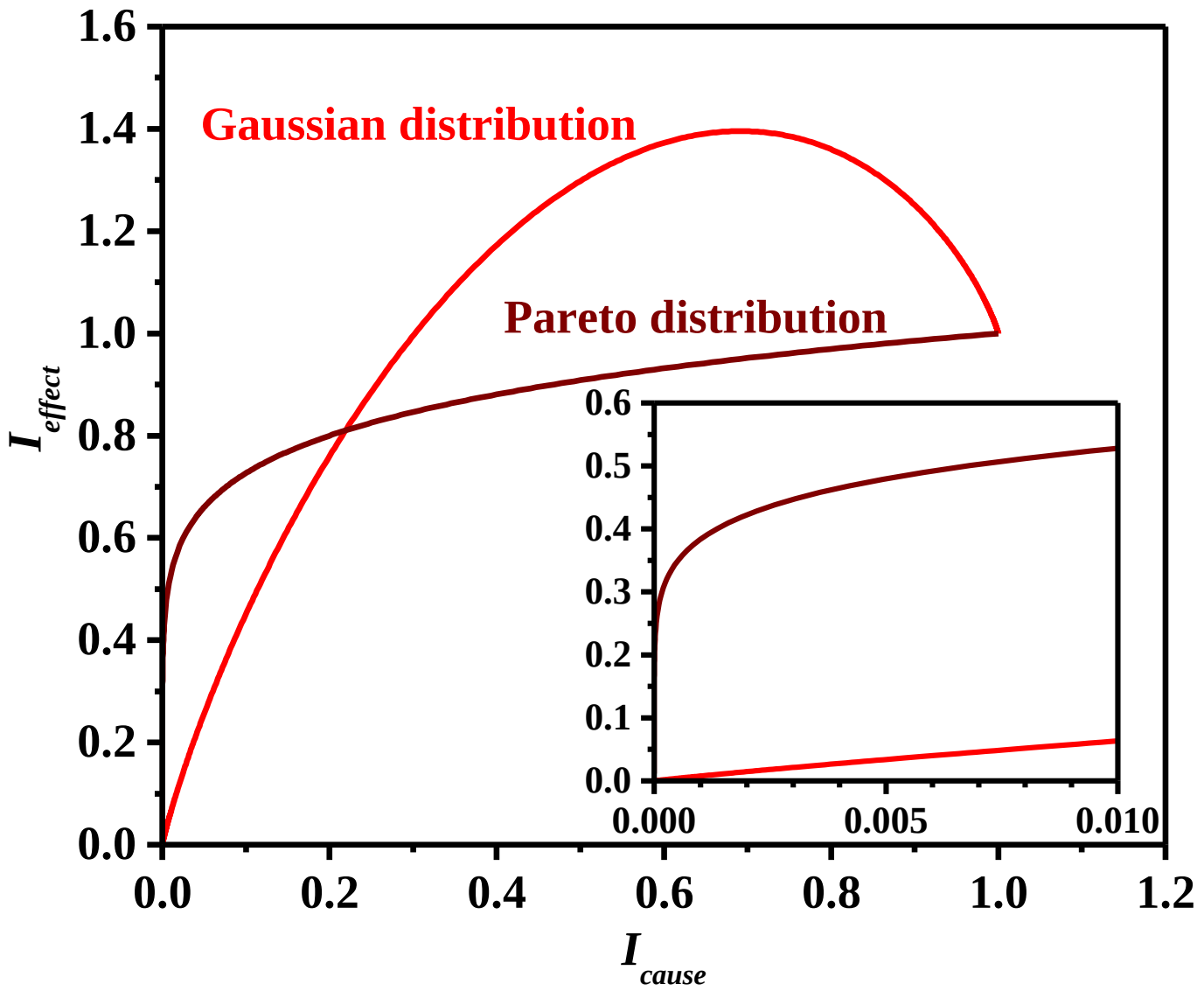

Fig. 5.